\documentclass[12pt]{article}
\usepackage{amsmath,amsfonts}

\begin{document}

\begin{center}
\textbf{Typical property of one class of combinatory objects and estimation from above corresponding combinatory numbers }

B.S. Kochkarev

{Kazan (Volga region) federal university, Russia}

E-mail: bkochkar@kpfu.ru
\end{center}
\vskip 12pt\noindent
{\bf Abstract}.\hskip 12pt {\sl  We investigate properties of families $F$ of subsets of a finite set in a situation where subsets are incomparable by the binary inclusion relation and a) for any $A\notin F$, there is such set $A'\in F$ that either $A\subset A'$ or $A'\subset A$; b) for any $A\in F$, $\mid A\mid\in \{k,k+1\}$. For these families we introduce one parametre and we show that for almost all families $F$ the value of this parametre is ${ n-1\choose k}$. We show that families with the minimum value of the entered parametre have certain structure and we find also number of such families. At last, we find an estimation from above for combinatory numbers of considered combinatory objects.}\vskip 12pt

Let $\Sigma(n)$ the variable set, which cardinality monotonically grows with the natural $n$.A property $\alpha$ concerning of elements of the set $\Sigma (n)$ is called as typical [1], if $\lim_{n \to \infty}\frac{\mid \Sigma_{\alpha}(n)\mid}{\mid \Sigma (n)\mid}=1$ (or $\lim_{n \to \infty}\frac{\mid \Sigma_{\overline{\alpha}}(n)\mid}{\mid \Sigma(n) \mid}=0$), where $\Sigma_{\alpha}(n)(\Sigma_{\overline{\alpha}}(n))$ signify set of elements from $\Sigma(n)$, which possess (do not possess) property $\alpha$.In this case sometimes say also that $\alpha$ it is fulfilled almost for all elements from $\Sigma(n)$.

We consider as $\Sigma(n)$ a class maximal Sperner families (m.S.f.) of subsets $F=F^{(k) }\cup F^{(k+1)}$ of the type $(k,k+1)$ [2-4] of finite ordered set $S=(a_{1},a_{2},\ldots,a_{n})$ and as property $\alpha$ the value of parametre $r(F)={ n-1 \choose k}$ [2-4] of the m.S.f. $F$ of the type $(k,k+1)$.

 Further we will speak about type only in special cases $k$. For $k$ we will consider [3,4] values $k<\lceil \frac{n}{2}\rceil, k\neq0$.

Definition 1 [5]. A family $F$ of subsets of the set $S$ is called Sperner, if no element $A\in F$ is a subset of another element $A'\in F$.

 Definition 2 [2-4]. A Sperner family (S.f.) $F$ is called maximal, if for any $A\subset S, A\notin F$, one can find $A'\in F$ such that $A\subset A'$ or $A'\subset A$.

 Definition 3 [ibid]. We say that a S.f. $F$ has the type $(k,k+1)$, if $\mid A\mid \in\{k,k+1\}$ for any $A\in F$.

Let $p_{i}$ stand for the number of elements $A\in F$, $\mid A\mid =k$, which do not contain the element $a_{i}\in S$; let $q_{i}$ stand for the number of elements $A\in F$, $\mid A\mid =k+1$ which contain the element $a_{i}$. Let $r_{i}=p_{i}+q_{i}, r=max r_{i}, i=\overline {1,n}$. Evidently, with any $n\geq 3$ the following inequality $r_{i}\leq { n-1 \choose k}$ is true.

 In [3] for the necessary condition received in [2] (theorem 2) it is made specification, is shown that it is true only for $n\leq 5$. For all $n\geq 6$ in [3] we have constructed m.S.f. with $r< { n-1 \choose k}$. Further we will consider $n\geq 6$.

 Theorem 1. If $F$ is a m.S.f., then $r(F)< { n-1 \choose k}$ if and only if for any $i= \overline {1,n}$ there exists a subset $A, \mid A\mid =k, a_{i} \notin A$ such that neither $A$, nor $A\cup \{a_{i}\}$ not belongs $F$.

Proof. Let $r(F)< { n-1 \choose k}$ and $i\in \overline {1,n}$. Let further $\tilde{F}\subset F$ is the family which consist of all $A, \mid A\mid=k, a_{i}\notin A$ and of all $B, \mid B\mid=k+1, a_{i}\in B$. Since $r(F)< { n-1 \choose k}$, then $\mid \tilde{ F}\mid< { n-1 \choose k}$. Hence, there exists a subset $A, \mid A\mid=k, a_{i}\notin A$, such that neither $A$, nor $A\cup \{a_{i}\}$ do not belong $F$.

 Let now for any $i\in \overline {1,n}$ it is possible to find $A, \mid A\mid =k, a_{i} \notin A$, such that neither $A$, nor $A\bigcup \{a_{i}\}$, do not belong $F$. From this follow that $r_{i}(F)< { n-1 \choose k}$. Consequently it is $r(F)< { n-1 \choose k}$.

 Corollary. $r(F)= { n-1 \choose k}$ if and only if there exists $i\in \overline {1,n}$ such that for any $A, \mid A\mid =k, a_{i}\notin A$, either $A\in F$ or $(A\bigcup \{a_{i}\})\in F$.

Theorem 2. For any the m.S.f. $F$ of type $(1,2)$ $r(F)= { n-1 \choose 1}$.

Proof. Really, let $F=F^{(1)}\bigcup F^{(2)}$ is a m.S.f.. Evidently, if $F=F^{(1)}=\{A, \mid A\mid=1\}$  or $F=F^{(2)}=\{B, \mid B\mid =2\}$, then $r_{i}(F)={ n-1 \choose 1}, i=\overline {1,n}$, thus is $r(F)={ n-1 \choose 1}$. We will admit now, that $F=F^{(1)}\bigcup F^{(2)}$, where $F^{(i)}\neq \emptyset, i=1,2$. Let $F^{(1)}=\{\{a_{i_{1}}\},\{a_{i_{2}}\},\ldots, \{a_{i_{k}}\}, k\neq n-1\}$. Then, evidently, $F^{(2)}=\{\{a_{j_{l}},a_{j_{m}}\}, 1\leq j_{l}<j_{m}\leq n, j_{l}\notin \{i_{1},i_{2},\ldots, i_{k}\}, j_{m}\notin \{i_{1},i_{2},\ldots,i_{k}\}\}$ and $r_{j_{i}}(F)= { n-1 \choose 1}$.

 Theorem 3. If $n\leq 5$, then for any m.S.f. $F$ $r(F)= { n-1 \choose k}$.

Proof. It can be checked up directly.

Theorem 4. If $F=F^{(k)}\bigcup F^{(k+1)}$ is the m.S.f. with $r_{n}(F)=0$, then $r_{i}(F)= { n-1 \choose k}, i\neq n$.

Proof. Really, for any $i\neq n$ the number of subsets $A, \mid A\mid =k, a_{i}\notin A$ is ${ n-2 \choose k-1}$ and the number of subsets $B, \mid B\mid =k+1, a_{i}\in B$ is ${ n-2 \choose k}$.From here $r_{i}(F)={ n-2 \choose k-1}+{ n-2 \choose k}={ n-1 \choose k}$.

                 Recursive algorithm of construction of all m.S.f.

For the proof of the basic result we use recursive algorithm of construction of all m.S.f.. If $F$ is a m.S.f., then $F'$ designates S.f. received after addition to $F$ a subset $A\notin F$, $A\subset S$ and exceptions from $F$ of all subsets $A'$ comparable with $A$ by inclusion. If $F$ is a S.f. such that for any $A\in F$ $a_{n}\in A$, then $F\setminus \{a_{n}\}$ designates S.f. which be formed out from $F$ by exception of all subsets $A\in F$ of the element $a_{n}$, and if $F$ is a S.f. such that for any $A\in F$ $a_{n}\notin A$, then $F\bigcup \{a_{n}\}$ designates the S.f., which be formed from $F$ by addition to each subset $A$ of the element $a_{n}$. It is exists only one the m.S.f. $F=F^{(k)}\bigcup F^{(k+1)}$ with $r_{n}(F)=0$, which consists of all the subsets $B, \mid B\mid =k+1, a_{n}\notin B$ and of all the subsets $A, \mid A\mid =k, a_{n}\in A$. Further the construction of the m.S.f. we organize recursively. Let $\Omega (t)$ is set of all m.S.f. constructed on a step $t$ $(\Omega (1)=\{F, r_{n}(F)=0\})$. The set $\Omega (t+1)$ is formed by means of transformations of the m.S.f. from $\Omega (t)$. Let $F=F^{(k)}\bigcup F^{(k+1)}$ is a m.S.f. from $\Omega (t)$. If $F$ is the m.S.f. with $r_{n}(F)={ n-1 \choose k}$, then we leave $F$ without change. If $F$ is a m.S.f. with $r_{n}(F)<{ n-1 \choose k}$ and

$M=\{A:\mid A\mid =k, a_{n}\notin A,A\notin F^{(k)}\}$,

$L=\{B:\mid B\mid =k+1, a_{n}\in B, B\notin F^{(k+1)}\}$,

$\Psi (F)=M\bigcup L\bigcup (M\bigcup \{a_{n}\})\bigcup (L\setminus \{a_{n}\})$,
then we carry out over $F$ the following transformations: we add to $F$ a subset from $\Psi (F)$ and we form family $F'$. After the exception from $F$ of all subset comparable by inclusion with the added subset can happen that among subsets in $\Psi (F)$ there are subsets incomparable by inclusion with subsets from $F'$. After addition of these subsets we will receive obviously the m.S.f.. As a result of these transformations with $F\in \Omega (t)$ with $r_{n}(F)<{ n-1 \choose k}$ concerning all subsets from $\Psi (F)$ we will receive set of the m.S.f., which form set $\Omega (t+1)$. Evidently $\bigcup _{t=1}^{ n-1 \choose k}$ is set of all m.S.f..

Theorem 5. Let $F=F^{(k)}\bigcup F^{(k+1)}, F^{(i)}\neq \emptyset, i\in \{k,k+1\}$ is the m.S.f. such that for any $B\in F^{(k+1)}$ $a_{i}\in B$ (for any $A\in F^{(k)} a_{i}\notin A$). Then $r(F)=r_{i}(F)= { n-1 \choose k}$.

Proof. Evidently, $q_{i}(F)< { n-1 \choose k}$ ($p_{i}(F)< { n-1 \choose k}$). From here follow that set $\{ A,\mid A\mid =k\}\setminus (F^{(k+1)}\setminus \{a_{i}\})\subseteq F^{(k)}$ ($\{ B,\mid B\mid =k+1\}\setminus (A\bigcup \{a_{i}\})\subseteq F^{(k+1)}$), that is $r_{i}(F)={ n-1 \choose k}$.

Corollary 1. For any the m.S.f. $F$ such that $\mid F^{(k+1)}\mid =1$ ($\mid F^{(k)}\mid =1$) $r(F)={ n-1 \choose k}$.

Corollary 2. If $n$ is odd, then for any m.S.f. $F$ of type $(\lfloor \frac {n}{2}\rfloor,\lceil \frac {n}{2}\rceil)$ such, that $\mid F^{(\lceil \frac {n}{2}\rceil)}\mid =2, r(F)={ n-1 \choose k}$

Lemma 1. Let $\hat{F}=\hat{F}^{(k)}\bigcup \hat{F}^{(k+1)}$ is the m.S.f. which received by the recursive algorithm from the m.S.f. $F=F^{(k)}\bigcup F^{(k+1)}$ with $r_{n}(F)<{ n-1 \choose k}$ in a step $t$. Then, if $r_{i}(F)={ n-1 \choose k}$ and $\hat{F}$ received by addition of subset $A, \mid A\mid =k, a_{i}\notin A$ ($B, \mid B\mid =k+1, a_{i}\in B$), then $r_{i}(\hat{F})={ n-1 \choose k}$.

Proof. Since $r_{i}(F)={ n-1 \choose k}$, then according to corollary to theorem 1 for any $A, \mid A\mid =k, a_{i}\notin A, A\notin F$ it is find $B, \mid B\mid=k+1, a_{i}\in B, B\supset A$, such that $B\in F$ (for any $B, \mid B\mid =k+1, a_{i}\in B, B\notin F$ it is find $A, \mid A\mid =k, a_{i}\notin A, A\subset B$ such that $A\in F$). Therefor, according to the recursive algorithm $r_{i}(\hat{F})=(p_{i}(F)+1)+(q_{i}(F)-1)={ n-1 \choose k}$ ($r_{i}(\hat{F})=(p_{i}(F)-1)+(q_{i}(F)+1)={ n-1 \choose k}$).

Lemma 2. If $F=F^{(k)}\bigcup F^{(k+1)},F^{(i)}\neq \emptyset, i\in \{k,k+1\}$ is the m.S.f. such, that $r(F)={ n-1 \choose k}=r_{i}(F), i\neq n$ and $r_{j}(F)\neq { n-1 \choose k}$ for $j\neq i$, then the m.S.f. $\tilde{F}$, which receive by the recursive algorithm by addition any subset $B, \mid B\mid =k+1, B\notin F, a_{i}\notin B$ $(A,\mid A\mid=k,A\notin F,a_{i} \in A)$ is the m.S.f. with $r(F)<{ n-1 \choose k}$.

Proof. Really, subset $B$ ($A$) comparable by inclusion with some subsets  $A$, $\mid A\mid =k, A\in F$ $(B,\mid B\mid =k+1,B\in F)$. Therefor, according to the recursive algorithm this subsets exception from $F$ and we receive $r_{i}(\tilde{F})={ n-1 \choose k}-m$, where $m$ is number of subsets exception from $F$; for $r_{j}(\tilde{F}), j\neq i$, evidently we will to have $r_{j}(\tilde{F})<{ n-1 \choose k}$.

Theorem 6. Let $F=F^{(k)}\bigcup F^{(k+1)}$, $F^{(j)}\neq \emptyset$ $j\in \{k,k+1\}$ is a m.S.f.. Then $r_{i}(F)={ n-1 \choose k}-\mid F_{i}^{(k)}\mid +\mid F_{i}^{(k+1)}\mid$, where $F_{i}^{(k+1)}$ is the family of all subsets $B$ from $F^{(k+1)}$ such that $a_{i}\in B$ and $F_{i}^{(k)}$ is the family of all subsets $A, \mid A\mid =k$, $a_{i}\notin A$ comparable with the subsets from $F^{(k+1)}$.

Proof. Really since ${ n-1 \choose k}-\mid F_{i}^{(k)}\mid =p_{i}(F)$ and $\mid F_{i}^{(k+1)}\mid =q_{i}(F)$, then $r_{i}(F)=p_{i}(F)+q_{i}(F)={ n-1 \choose k}-\mid F_{i}^{(k)}\mid +\mid F_{i}^{(k+1)}\mid$.

Corollary 1. If $F=F^{(k)}\bigcup F^{(k+1)}$, $F^{(j)}\neq \emptyset, j\in\{k,k+1\}$ is m.S.f. such that $max_{r}q_{r}(F)=q_{i}(F)$ then $max_{r}p_{r}(F)=p_{i}(F)$ and $r(F)=r_{i}(F)$.

Proof. Really, according to theorem 6 $p_{j}(F)={ n-1 \choose k}-\mid F_{j}^{(k)}\mid$, where $F_{j}^{(k)}$ is the family of all subsets $A, \mid A\mid =k, a_{j} \notin A$ comparable with the subsets from $F^{(k+1)}$, but since $\mid F_{j}^{(k)}\mid\geq \mid F_{i}^{(k)}\mid$ then $p_{j}(F)\leq p_{i}(F)$ and consequently $max_{j}p_{j}(F)=p_{i}(F), r(F)=r_{i}(F)$.

Corollary 2. If $F=F^{(k)}\bigcup F^{(k+1)}, F^{(i)}\neq \emptyset, i\in \{k,k+1\}$ is m.S.f. such that for any pair $B_{i},B_{j}$ of subsets from $F^{(k+1)}$ $B_{i}\bigcap B_{j}=\emptyset$ and $\sum_{B_{i}\in F^{(k+1)}}$ $\mid B_{i}\mid=(k+1)\mid F^{(k+1)}\mid, (k\neq 1)$, then $r_{i}=r(F)$, if $a_{i}\in B$ for a subset $B$ from $F^{(k+1)}$.

Proof. Really, according to theorem 6 we have $r_{i}={ n-1 \choose k}-\mid F^{(i)}\mid +1$, if $a_{i}$ belong to a $B$ from $F^{(k+1)}$ and $r_{j}(F)<r_{i}(F)$ if $a_{i}$ do not belong to none $B$ from $F^{(k+1)}$, i.e. $r(F)=r_{i}(F)$. Evidently, if $\frac {n}{k+1}$ is the whole and $\sum_{B_{i}\in F^{(k+1)}}\mid B_{i}\mid=n$ then for all $i=\overline {1,n}$ $r(F)=r_{i}(F)={ n-1 \choose k}-\mid F_{i}^{(k)}\mid +1={ n-1 \choose k}-(\frac{n}{k+1}-1)(k+1)$.

Corollary 3. If $k<\lfloor \frac{n}{2}\rfloor, F=F^{(k)}\bigcup F^{(k+1)}, F^{(i)}\neq \emptyset, i\in\{k,k+1\}$ is m.S.f. such that for any pair $B_{i},B_{j}$ of subsets from $F^{(k+1)}$ $B_{i}\bigcap B_{j}=\emptyset$ and $\mid F^{(k+1)}\mid=\lfloor \frac{n}{k+1}\rfloor$, then $r(F)=min_{F:r(F)<{ n-1 \choose k}}r(F)$ and the number of such m.S.f. is

${ n \choose k+1}{ n-(k+1) \choose k+1}\cdots { n-(\lfloor \frac{n}{k+1}\rfloor-1) \choose k+1}$.

Proof. According to corollary 2 $r(F)=r_{i}(F)={ n-1 \choose k}-\mid F_{i}^{(k)}\mid +1$, if $a_{i}$ belong to a $B$ from $F^{(k+1)}$. Evidently in the conditions of corollary 3 $\mid F_{i}^{(k)}\mid=max_{F:r(F)<{ n-1 \choose k}}\mid F_{i}^{(k)}\mid$. The second affirmation evidently. Evidently also, that ${ n-1 \choose k}-(\lfloor \frac{n}{k+1}\rfloor-1)(k+1)\leq r(F)\leq { n-1 \choose k},r(F)\neq { n-1 \choose k}-1$.

Corollary 4. If $n$ is even, then the m.S.f. $F=F^{(\frac {n}{2}-1)}\bigcup F^{(\frac {n}{2})}$, where $F^{(\frac {n}{2})}=\{\{a_{i_{1}}, a_{i_{2}}, \ldots, a_{i_{\frac {n}{2}}}\},\{a_{j_{1}}, a_{j_{2}}, \ldots, a_{j_{\frac {n}{2}}}\}:i_{k}\notin \{j_{1}, j_{2}, \ldots, j_{\frac {n}{2}}\}, k=\overline {1,\frac {n}{2}}\}$ is the m.S.f. with $r(F)=r_{i}(F)={ n-1 \choose \frac {n}{2}-1}-\frac {n}{2}, i=\overline {1,n}$. Evidently the number of these m.S.f. is ${ n \choose \frac{n}{2}}2^{-1}$.

Theorem 5 delivery to us the sufficient condition for $r(F)=r_{i}(F)={ n-1 \choose k}$. We will prove that this condition there is also the necessary condition.

Corollary 5. If $r(F)=r_{i}(F)={ n-1 \choose k}$, then any $B\in F^{(k+1)}$ $a_{i}\in B$.

Proof. Really, according to theorem 6 $r_{i}(F)={ n-1 \choose k}-\mid F_{i}^{(k)}\mid$ $+\\
\mid F_{i}^{(k+1)}\mid$, but since $r_{i}(F)={ n-1 \choose k}$, then $\mid F_{i}^{(k+1)}\mid$ $=\mid F_{i}^{(k)}\mid$, consequently for all $B\in F^{(k+1)}$ $a_{i}\in B$.

Theorem 7. If $n$ is odd and $F=F^{(\lfloor \frac{n}{2}\rfloor)} \bigcup F^{(\lceil \frac{n}{2}\rceil)}$ is m.S.f., then minimum
$\mid F^{(\lceil \frac{n}{2}\rceil)}\mid$ such that $r(F)<{ n-1 \choose \lfloor \frac{n}{2}\rfloor}$ is $3$.

Proof.  According to corollary 2 to theorem 5, if $n$ is odd, then for any m.S.f. $F$ of
type $(\lfloor \frac{n}{2}\rfloor,\lceil \frac{n}{2}\rceil)$ such,
that $\mid F^{(\lceil \frac{n}{2}\rceil)}\mid=2$ $r(F)={ n-1 \choose \lfloor \frac{n}{2}\rfloor}$.
Let now $F=F^{(\lfloor \frac{n}{2}\rfloor)}\bigcup F^{(\lceil \frac{n}{2}\rceil)}$ be the m.S.f.,
where $F^{(\lceil \frac{n}{2}\rceil)}=\{B_{1},B_{2},B_{3}: B_{1}\bigcap B_{2}=a_{i}\neq a_{n}; a_{i}\notin B_{3}\}$. Evidently the number of these m.S.f. is $(n-1){ n-2 \choose \lfloor \frac {n}{2}\rfloor}{ n-1 \choose \lceil \frac{n}{2}\rceil}$. If the subset $B_{3}$ not have of the common comparable subset $A,\mid A\mid=\lfloor \frac{n}{2}\rfloor$ with the subsets $B_{1},B_{2}$,then according to theorem 6 $r(F)={ n-1 \choose \lfloor \frac{n}{2}\rfloor}-\lceil \frac{n}{2}\rceil$ and $min_{i}r_{i}(F)={ n-1 \choose \lfloor \frac{n}{2}\rfloor}-2\lceil \frac{n}{2}\rceil$, but if the subset $B_{3}$ has the common comparable subset $A, \mid A\mid=\lfloor \frac{n}{2}\rfloor$, with a subset from $\{B_{1},B_{2}\}$, then $r(F)={ n-1 \choose k}-\lceil \frac{n}{2}\rceil+1$ and $min_{i}r_{i}(F)={ n-1 \choose k}-2\lceil \frac{n}{2}\rceil+1$.

Theorem 8. Almost for all m.S.f. $F$ $r(F)={ n-1 \choose k}$.

Proof. We will prove this only for the m.S.f. of the type $(\lfloor \frac{n}{2}\rfloor,\lceil \frac
{n}{2}\rceil)$ with $n$ odd since for other cases the proof carry out analogously.We will estimate
 from above the number of m.S.f. $F$ such that $r(F)<{ n-1 \choose \lfloor \frac{n}{2}\rfloor}$.
 Evidently, according to the recursive algorithm and to theorem 4 the number of these m.S.f. not
  exceed of the number of m.S.f. $F$ with $r(F)<{ n-1 \choose \lfloor \frac{n}{2}\rfloor}$ received
   before m.S.f. $F$ with $min_{F:r(F)<{ n-1 \choose \lfloor \frac{n}{2}\rfloor}}r(F)$ plus of the
    number of m.S.f. $F'$ with $r_{n}(F')<{ n-1 \choose \frac{n}{2}}$ received later the m.S.f. $F$
     with $min_{F:r(F)<{ n-1 \choose \lfloor \frac{n}{2}\rfloor}}r(F)$. According to lemma 2 and to the recursive algorithm the number $\mid \sum_{\bar{\alpha}}(n)\mid$  m.S.f. with $r(F)<{ n-1 \choose \lfloor \frac{n}{2}\rfloor}$ satisfy to inequality

     $\mid \sum_{\bar{\alpha}}(n)\mid\leq (n-1)^{2}{ n-2 \choose \lfloor \frac{n}{2}\rfloor}{ n-1
      \choose \lceil \frac{n}{2}\rceil}\lceil\frac{n}{2}\rceil + 2\lceil \frac{n}{2}\rceil(n-1)
      { n-2 \choose \lfloor \frac{n}{2}\rfloor}{ n-1 \choose \lceil \frac{n}{2}\rceil}({ 2\lceil
       \frac{n}{2}\rceil \choose 0}+{ 2\lceil \frac{n}{2}\rceil \choose 1}2^{1}+\ldots+{ 2\lceil
        \frac{n}{2}\rceil \choose \lceil \frac{n}{2}\rceil-2}2^{2\lceil \frac{n}{2}\rceil-2})<2\lceil \frac{n}{2}\rceil(n-1){ n-2 \choose \lfloor \frac{n}{2}\rfloor}{ n-1 \choose \lceil \frac{n}{2}\rceil}2^{4\lceil \frac{n}{2}\rceil-2}$. Since the number of m.S.f. with $r(F)={ n-1 \choose \lfloor \frac{n}{2}\rfloor}$ more than $2^{{ n-1 \choose \lfloor \frac{n}{2}\rfloor}}$ then
        $\lim \limits_{n \to \infty}\frac{\mid \sum_{\bar{\alpha}}(n)\mid}{\mid \sum(n)\mid} =0$. From here directly follow $\mid \sum(n)\mid<n2^{{ n-1 \choose k}}$.

        The above estimate received for all m.S.f. improve essentially the estimate $2^{3{ n-1 \choose k}}$ from [3].

        Theorem 9. For $k=1$ $\mid\sum(n)\mid\sim 2^{n}=2\cdot2^{ n-1 \choose 1}$.

        Proof. It is trivial to show that in this case $\mid \sum(n)\mid=2^{n}-n$. Thus in this case we have $\mid \sum(n)\mid\sim 2^{n}=2\cdot2^{ n-1 \choose 1}$.

        Hypothesis: we suppose that for any $k$ $\mid\sum(n)\mid\sim (k+1)2^{ n-1 \choose k}$.

\vskip5mm
\centerline{REFERENCES}
\smallskip

1. A.A. Sapogenko, Metricheskie svoystva pochti vsekh funktsiy algebry logiki, Diskretny analiz, vyp. 10, Novosibirsk, IM SO AN SSSR, 1967, pp. 91-119.

2. B.S. Kochkarev, Structural properties of a class of maximal Sperner families of the finite set, Logika i prilogeniya, Tez. Megdunar. konf. posv. 60-letiju so dnya rogd. acad. Ju.L. Ershova, Novosibirsk, 2000, pp. 61-62.

3. B.S. Kochkarev, Structural properties of a class of maximal Sperner families of subsets, Izv.Vyssh. Uchebn. Zaved. Mat, 2005, ¹7, pp. 37-42.

4. B.S. Kochkarev, Admissible Values of One Parameter for Maximal Sperner Families of Subsets of the Type (k,k+1), Russian Mathematics (Iz.VUZ),2008, Vol. 52, ¹6, pp.22-24.

5. E. Sperner, Ein Satz $\ddot{ u}ber$ Untermengen einer edlichen Menge, Math. Z. 27 (1928), 544-548.

\end{document}